\documentclass[aps,pra,twocolumn,superscriptaddress]{revtex4}
\usepackage{amsfonts,amssymb,graphicx}
\usepackage{amsmath}
\usepackage[usenames,dvipsnames]{xcolor}
\usepackage{stackrel}
\usepackage[normalem]{ulem}
\usepackage{color}
\begin{document}

\title{Collective charging of an organic quantum battery}
\author{Jiawei Li}
\affiliation{Center for Quantum Technology Research, and Key Laboratory of Advanced Optoelectronic Quantum Architecture and Measurements (MOE), School of Physics, Beijing Institute of Technology, Beijing 100081, China}
\author{Ning Wu}
\email{wunwyz@gmail.com}
\affiliation{Center for Quantum Technology Research, and Key Laboratory of Advanced Optoelectronic Quantum Architecture and Measurements (MOE), School of Physics, Beijing Institute of Technology, Beijing 100081, China}
\begin{abstract}
We study the collective charging of a quantum battery (QB) consisting of a one-dimensional molecular aggregate and a coupled single-mode cavity, to which we refer as an ``organic quantum battery" since the battery part is an organic material. The organic QB can be viewed as an extension of the so-called Dicke QB [D. Ferraro \emph{et al}., Phys. Rev. Lett. \textbf{120}, 117702 (2018)] by including finite exciton hopping and exciton-exciton interaction within the battery. We consider two types of normalizations of the exciton-cavity coupling when the size of the aggregate $N$ is increased: (I) The cavity length also increases to keep the density of monomers constant, (II) The cavity length does not change. Our main findings are that: (i) For fixed $N$ and exciton-cavity coupling, there exist optimal exciton-exciton interactions at which the maximum stored energy density and the maximum charging power density reach their respective maxima that both increase with increasing exciton-cavity coupling. The existence of such maxima for weak exciton-cavity coupling is argued to be due to the non-monotonic behavior of the one-exciton to two-exciton transition probability in the framework of second-order time-dependent perturbation theory. (ii) Under normalization I, no quantum advantage is observed in the scaling of the two quantities with varying $N$. Under normalization II, it is found that both the maximum stored energy density and the maximum charging power density exhibit quantum advantages compared with the Dicke QB.
\end{abstract}

\maketitle
\section{Introduction}
\par Recently, there has been a growing interest in the study of the so-called quantum batteries (QBs)~\cite{RMP}, which are energy storage systems that utilize quantum mechanical objects. A variety of models of many-body QBs have been theoretically proposed, including the Dicke QB~\cite{QB2018}, the spin-network~\cite{Guan2021,Yang2021} and spin-chain batteries~\cite{Le2018,Grazi2024}, the SYK batteries~\cite{SYK}, and the bosonic batteries~\cite{bosonQB}, etc. Among these, the Dicke QB~\cite{QB2018} inspired by the superradiance phenomenon~\cite{Dicke} was proposed with the hope of achieving a superextensive scaling in the charging power. The Dicke QB consists of $N$ noninteracting two-level systems (the battery part) and a single-mode cavity field (the charger). The spin-network or spin-chain QBs serve as another interesting example of a QB, in which the charging protocol is realized by tuning self-interactions among the spins or via direction charging protocols.
\par In this work, we study the charging process of a QB that can be viewed as a combination of the aforementioned two types of QBs. We consider a generalization of the Dicke QB to the case where the noninteracting two-level atoms are replaced by an organic molecular aggregate modeled by an interacting Frenkel exciton model~\cite{Ezaki1994,SpanoCPL1995,JCP1997}. With the exciton hopping and exciton-exciton interaction included, energy can be stored in the interaction part of the molecular system. Such a composite system can be experimentally realized by locating a molecular aggregate in a micocavity and has been widely studied in the field of strong light-matter couplings in the last decade~\cite{Spano2015,LPP2016,JCP2019,JCP2020}. We thus refer to our battery as an organic QB because of the organic nature of the battery part. Since real-space Frenkel excitons are equivalent to hardcore bosons that behave like bosons on distinct sites while like fermions on a single site, the molecular battery is indeed described by a spin-1/2 XXZ chain~\cite{Ezaki1994,SpanoCPL1995,JCP1997}. Unlike the pure Dicke QB in which the total angular momentum of the pseudo-spins is conserved~\cite{QB2018}, this may gives rise to difficulties in simulating the real-time dynamics of the QB since all angular momentum sectors are involved in the charging process~\cite{XXZCSM}. To this end, we resort to an exact diagonalization method based on the spin-operator matrix elements~\cite{XXZCSM,Wu2018} to simulate the real-dynamics of the system for aggregates of $N\leq 18$ monomers.
\par To study the scaling behavior of the maximum stored energy density and maximum charging power density with increasing aggregate size, two different types of normalization procedures of the exciton-cavity coupling are adopted. Under the first type of normalization (type I), the cavity length proportionally increases with the enlarged aggregate and the system admits a well-defined thermodynamic limit~\cite{LPP2016,PRR2020}. Under the other type of normalization (type II), the cavity length is kept constant when $N$ increases~\cite{QB2018}. By simulating the charging process for a variety sets of parameters, we find that generally no quantum advantage shows up in both the maximum stored energy density and the maximum charging power density if normalization I is used. However, under normalization II, finite exciton hopping and exciton-exciton interactions could give rise to improved quantum advantages in the two quantities compared with the case of a Dicke QB~\cite{QB2018}. Interestingly, for fixed aggregate size and fixed exciton-cavity coupling, we observe optimal exciton-exciton interactions at which the two quantities reach their maximal values. We argue that the existence of these optimal exciton-exciton interactions arises from the non-monotonic behavior of the one-exciton to two-exciton probability as the exciton-exciton interaction increases.
\par The rest of the paper is organized as follows. In Sec.~\ref{SecII}, we introduce our model for an organic QB and provide the numerical method that is used to simulate the real-time dynamics of the battery. In Sec.~\ref{SecIII}, we present detailed numerical results on the charging process of the organic QB. Conclusions are drawn in Sec.~\ref{SecV}.
\section{Model and methodology}\label{SecII}
\subsection{The organic quantum battery and charging protocol}
\par
\begin{figure}
\includegraphics[width=.48\textwidth]{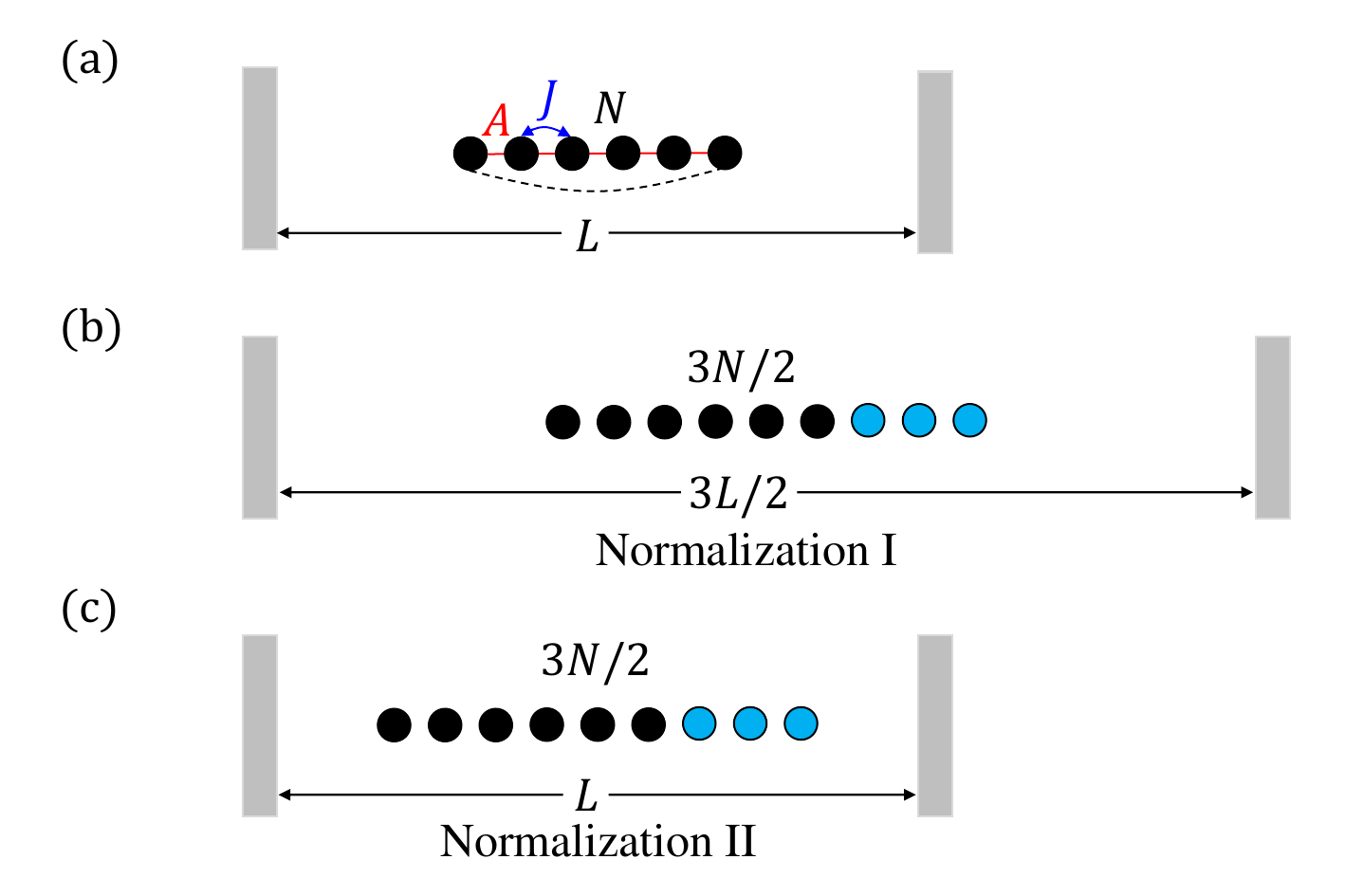}
\caption{(a) An organic quantum battery is described by an interacting Frenkel-Dicke model, where a molecular aggregate (the battery part) with nearest-neighbor dipole-dipole coupling $J$ and exciton-exciton interaction $A$ is coupled to a single cavity mode (the charger part). We use periodic boundary conditions (dashed) for simplicity. (b) Under normalization I, the cavity length $L$ increases proportionally with the aggregate length $N$ to keep the monomer density $N/L$ constant. (c) Under normalization II, the cavity length is unchanged when the aggregate length is increased.}
\label{Fig1}
\end{figure}

 We consider a one-dimensional molecular aggregate interacting with a single cavity mode via time-dependent light-matter interaction [Fig.~\ref{Fig1}(a)]. Such a composite system can be described by the following interacting Frenkel-Dicke model ($\hbar=1$)
\begin{eqnarray}
H_{\mathrm{F-D}}&=&H_{\mathrm{m}}+H_{\mathrm{c}}+H_{\mathrm{m-c}},\nonumber\\
H_{\mathrm{m}}&=&\omega\sum^N_{j=1}a^\dag_ja_j+J\sum^N_{j=1}(a^\dag_j a_{j+1}+a^\dag_{j+1}a_j)\nonumber\\
&&+A\sum^N_{j=1}a^\dag_j a_j a^\dag_{j+1}a_{j+1},\nonumber\\
H_{\mathrm{c}}&=&\omega_{\mathrm{c}}c^\dag c,\nonumber\\
H_{\mathrm{m-c}}&=&\sum^N_{j=1}g_j(t)(a^\dag_jc+a_j c^\dag).
\end{eqnarray}
\par Here, $a^\dag_j$ creates a Frenkel exciton on site $j$ with uniform excitation energy $\omega>0$, $J$ and $A$ are the homogenous nearest-neighbor dipole-dipole coupling and exciton-exciton interaction, respectively. The explicit expressions of $J$ and $A$ are given by~\cite{Ezaki1994,JPCL2014}
\begin{eqnarray}\label{J}
J&=&\frac{1}{4\pi\epsilon_0 d^3}\left(\mu^2-\frac{3(\vec{\mu}\cdot\vec{d})^2}{d^2}\right),\nonumber\\
A&=&\frac{1}{4\pi\epsilon_0 d^3}\left(\tilde{\mu}^2-\frac{3(\vec{\tilde{\mu}}\cdot\vec{d})^2}{d^2}\right),
\end{eqnarray}
where $\vec{d}$ is the relative position between nearest-neighbor monomers, $\vec{\mu}$ is the transition dipole moment of the monomer, and $\vec{\tilde{\mu}}$ is the static dipole moment induced by the excitation of the monomer. We assume periodic boundary conditions for the molecular aggregate, i.e., $a_{N+1}=a_1$. The interacting Frenkel-exciton model $H_{\mathrm{m}}$ has been previously employed in the study of multi-exciton physics and nonlinear optical response of molecular aggregates~\cite{Ezaki1994,SpanoCPL1995,JCP1997}.
\par $H_{\mathrm{c}}$ describes the single-mode cavity with photon creation operator $c^\dag$ and cavity frequency $\omega_{\mathrm{c}}$. $H_{\mathrm{m-c}}$ represents the exciton-cavity coupling under the rotating wave approximation (RWA), with time-dependent strength $g_j(t)$. We further assume that the size of the chain is much smaller than the optical wavelength so that the exciton-cavity coupling is nearly uniform, $g_j(t)=g(t)$. This approximation has been used by several authors in the study of molecular systems coupled to a light field~\cite{Spano2015,LPP2016,Nitzan2022}. The Frenkel-Dicke model under the single-photon approximation and in the absence of exciton-exciton interaction has been used to study exciton transport in molecular crystals strongly coupled to a cavity~\cite{LPP2016,JCP2019,JCP2020,JF2015}.
\par We are interested in the charging process of the molecular aggregate by the cavity photons. The total system is prepared in the product state
\begin{eqnarray}\label{psi0}
|\psi(0)\rangle=|n_{\mathrm{ph}}\rangle\otimes|G\rangle,
\end{eqnarray}
where $|n_{\mathrm{ph}}\rangle$ is the Fock state having $n_{\mathrm{ph}}$ photons and $|G\rangle$ is the ground state of $H_{\mathrm{m}}$ with energy $E_{G}$. The exciton-cavity coupling is turned on at time $t=0^+$, $g(t=0^+)=g$, and is kept at this value before the interaction is turned off when the storage step begins.
\par The coupling constant $g$ in $H_{\mathrm{m-c}}$ has its origin in the dipole coupling $-\vec{\mu}\cdot\vec{E}$, where $\vec{E}$ is the electric field at the aggregate. In a cavity of length $L$, the field $\vec{E}$ carries a normalization factor $1/\sqrt{L}$. In this work, we will consider two scenarios when the size of the molecular aggregate $N$ is changed: (I) The cavity length $L$ increases proportionally with increasing $N$ to keep the monomer density $N/L$ constant [see Fig.~\ref{Fig1}(b)]~\cite{LPP2016,PRR2020}. This normalization can lead to a well-defined thermodynamic limit and amounts to setting $g\sqrt{N}/\omega$ unchanged when $N$ varies~\cite{Hepp,Wang}. In addition, this kind of normalization can ensure that the short-time dynamics is universal for different $N$'s (see Fig.~\ref{N14vs16} below)~\cite{PRB2016}. (II) The cavity length $L$ is kept unchanged when the molecular aggregate is enlarged [see Fig.~\ref{Fig1}(c)], which means that $g$ is a constant when $N$ varies~\cite{QB2018}. Normalization II is adopted in Ref.~\cite{QB2018} to study the charging process of a conventional Dicke QB and a quantum advantage in the charging power that scales like $\sqrt{N}$ is demonstrated. In this work, we will consider both types of normalizations.
\par The energy stored in the molecular aggregate at time $t$ is defined as~\cite{QB2018}.
\begin{eqnarray}
E_{\mathrm{m}}(t)=\langle\psi(t)|H_{\mathrm{m}}|\psi(t)\rangle-E_G,
\end{eqnarray}
where $|\psi(t)\rangle=e^{-iH_{\mathrm{F-D}}(g)t}|\psi(0)\rangle$ is the time-evolved state during the charging process. The charing power is defined as~\cite{QB2018}
\begin{eqnarray}
P(t)=E_{\mathrm{m}}(t)/t.
\end{eqnarray}
\subsection{Method: spin-operator matrix elements}
\par As mentioned early, the Frenkel excitons are neither bosons nor fermions but are hardcore bosons~\cite{Spano1991}. This means that the exciton creation operator $a^\dag_j$ is actually a Pauli raising operator $a^\dag_j=S^x_j+iS^y_j$ ($S^\alpha_j$ is the $\alpha$-component of the spin-1/2 operator $\vec{S}_j$) and the molecular Hamiltonian $H_{\mathrm{m}}$ resembles a spin-1/2 XXZ spin chain~\cite{Ezaki1994,SpanoCPL1995}:
\begin{eqnarray}\label{XXZ}
H_{\mathrm{m}}&=&(\omega+A)\sum^N_{j=1}S^z_j+2J\sum^N_{j=1}(S^x_j S^x_{j+1}+S^y_{j+1}S^y_j)\nonumber\\
&&+A\sum^N_{j=1}S^z_jS^z_{j+1}+\frac{N}{4}(A+2\omega).
\end{eqnarray}
\par As a strongly correlated model, the real-time dynamics involving the nearest-neighbor XXZ chain is usually difficult to treat~\cite{XXZCSM} due to the lack of conserved quantities such as the total angular momentum~\cite{QB2018,Dou,Zhang}. Although the statistics of the Frenkel excitons is irrelevant in the single-excitation limit~\cite{JF2015,LPP2016,JCP2019,JCP2020}, it plays a nontrivial role when multi-exciton states are relevant. Here, we will simulate the dynamics of the organic QB in a basis that diagonalizes the noninteracting Frenkel exciton chain with $A=0$. The fundamental excitations in a noninteracting Frenkel exciton chain are known to be spinless fermions~\cite{Spano1991,Wu2018}:
\begin{eqnarray}
H^{(A=0)}_{\mathrm{m}}|\vec{\eta}_n\rangle=\mathcal{E}^{(A=0)}_{\vec{\eta}_n}|\vec{\eta}_n\rangle,
\end{eqnarray}
where $|\vec{\eta}_n\rangle$ is an eigenstate of $H^{(A=0)}_{\mathrm{m}}$ having $n$ fermionic excitations (with respect to the excitonic vacuum $|\mathrm{vac}\rangle=|\downarrow\ldots\downarrow\rangle$) labelled by the ordered $n$-tuple $\vec{\eta}_n=(\eta_1,\ldots,\eta_n)$ with $1\leq \eta_1<\cdots<\eta_n\leq N$ and $\mathcal{E}^{(A=0)}_{\vec{\eta}_n}=\omega+2J\sum^n_{l=1}\cos K^{(\sigma_n)}_{\eta_l}$ the corresponding eigenenergy. The wave numbers appearing in $\mathcal{E}^{(A=0)}_{\vec{\eta}_n}$ take values $K^{(\sigma_n)}_{\eta_l}=-\pi+[2\eta_l+\frac{1}{2}(\sigma_n-3)]\frac{\pi}{N}$ for even $N$ ($K^{(\sigma_n)}_{\eta_l}=-\pi+[2\eta_l-\frac{1}{2}(\sigma_n+3)]\frac{\pi}{N}$ for odd $N$), where $\sigma_n=1$ ($\sigma_n=-1$) if $n$ is even (odd).
\par The Hamiltonian $H_{\mathrm{F-D}}$ conserves the total number of excitations $\hat{\mathcal{N}}=\sum_j a^\dag_j a_j+c^\dag c$. As a consequence, for the initial state given by Eq.~(\ref{psi0}), the time evolution takes place in the subspace with $\mathcal{N}=n_{\mathrm{ph}}+n_{\mathrm{ex}}$, where $0\leq n_{\mathrm{ex}}\leq N$ is the number of excitations upon the ground state $|G\rangle$ of the molecular aggregate and depends on the parameters $J/\omega$ and $A/\omega$. For $\mathcal{N}\geq N$, this $\mathcal{N}$-subspace is spanned by the following basis states,
\begin{eqnarray}\label{basis1}
|\mathcal{N}\rangle|\mathrm{vac}\rangle,\ldots,|\mathcal{N}-m\rangle\{|\vec{\eta}_{m}\rangle\},\ldots,|\mathcal{N}-N\rangle|\vec{\eta}_{N}\rangle,
\end{eqnarray}
where $\{|\vec{\eta}_{m}\rangle\}$ represents the set formed by all the $\binom{N}{m}=\frac{N!}{m!(N-m)!}$ molecular basis states having $m$ excitons.
\begin{figure}
\includegraphics[width=.51\textwidth]{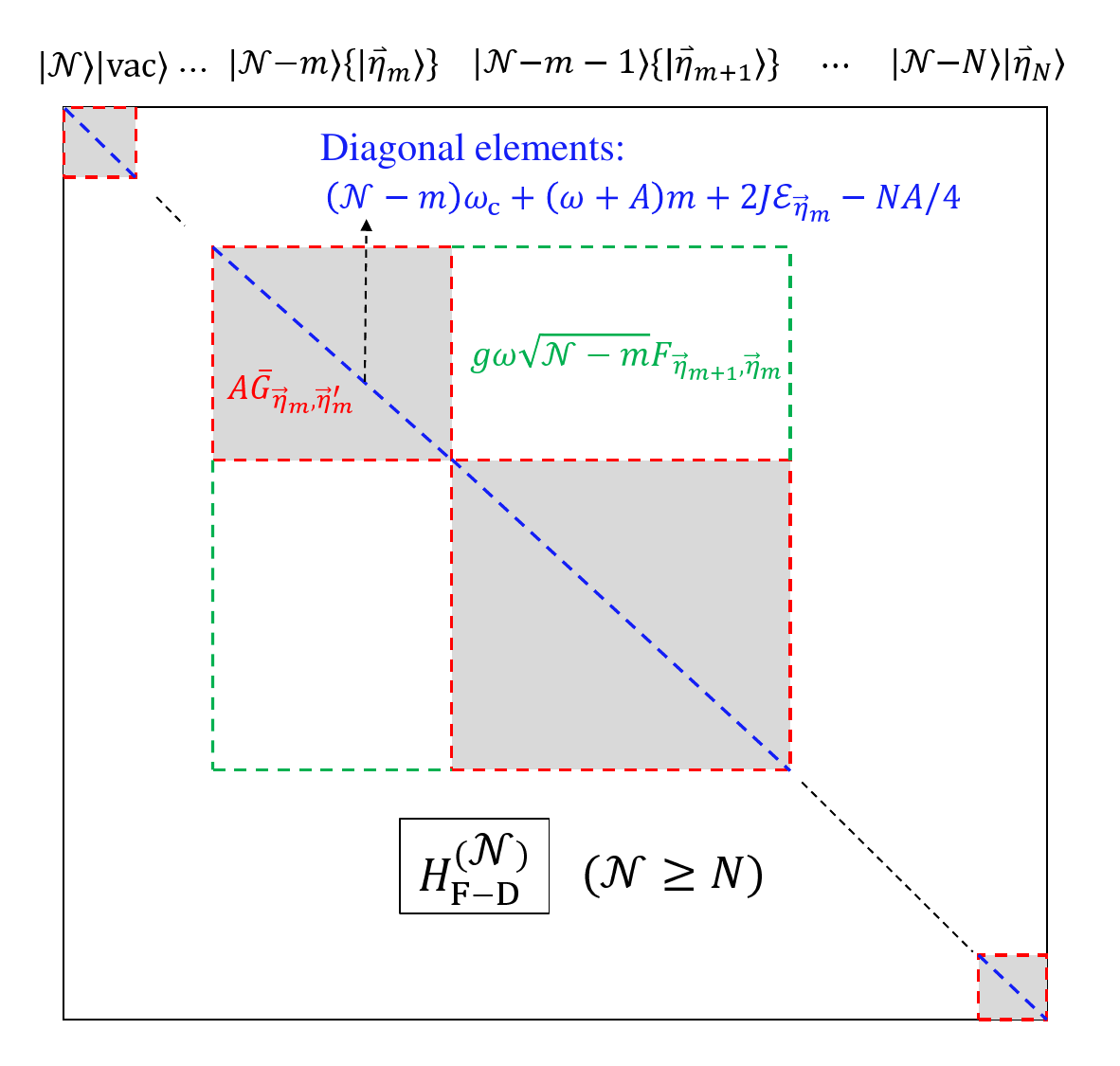}
\caption{Matrix representation of $H_{\mathrm{F-D}}$ in the subspace specified by the total excitation number $\mathcal{N}$ and spanned by the basis states given by Eq.~(\ref{basis1}).}
\label{Fig2}
\end{figure}
For $\mathcal{N}<N$, the $\mathcal{N}$-subspace is spanned by
\begin{eqnarray}\label{basis2}
|\mathcal{N}\rangle|\mathrm{vac}\rangle,\ldots,|\mathcal{N}-m\rangle\{|\vec{\eta}_{m}\rangle\},\ldots,|0\rangle|\vec{\eta}_{\mathcal{N}}\rangle.
\end{eqnarray}
Note that the dimension of the $\mathcal{N}$-subspace with $\mathcal{N}\geq N$ ($\mathcal{N}< N$) is $2^N$ [$\sum^{\mathcal{N}}_{m=0}\binom{N}{m}$].
\par The matrix representation $H^{(\mathcal{N})}_{\mathrm{F-D}}$ of the total Hamiltonian in the basis given by Eq.~(\ref{basis1}) has the form shown in Fig.~\ref{Fig2}, where explicit forms of the $F$- and $\bar{G}$-functions are~\cite{XXZCSM,Wu2018}
\begin{eqnarray}\label{F}
F_{\vec{\eta}_{m+1},\vec{\eta}_m}&=&\langle\vec{\eta}_m|\sum^N_{j=1}S^-_j|\vec{\eta}_{m+1}\rangle\nonumber\\
&=&2^mN^{\frac{1}{2}-m} \delta(\Delta_{\vec{\eta}_{m+1},\vec{\eta}_m},0)h_{\vec{\eta}_{m+1},\vec{\eta}_m}
\end{eqnarray}
and
\begin{eqnarray}\label{G}
\bar{G}_{\vec{\eta}_m,\vec{\eta}'_m}&=&\langle\vec{\eta}_m|\sum^N_{j=1}S^z_jS^z_{j+1}|\vec{\eta}'_{m}\rangle\nonumber\\
&=&\left(m -\frac{3 N}{4}\right)\delta_{\vec{\eta}_m,\vec{\eta}'_m}+\left(\frac{2}{N}\right)^{4m}\frac{\delta(\Delta_{\vec{\eta}_m,\vec{\eta}'_m},0)}{N}\nonumber\\
&&\times\sum_{\vec{\chi}_m}e^{ i\Delta_{\vec{\eta}_m,\vec{\chi}_m}} \bar{h}_{\vec{\eta}_m,\vec{\chi}_m}\bar{h}_{\vec{\chi}_m,\vec{\eta}'_m}.
\end{eqnarray}
Here,
\begin{eqnarray}
\delta(x,y)=\begin{cases}
    1,  & \text{$x-y=2\pi m,~m\in Z$},\\
    0, & \mathrm{otherwise},
  \end{cases}
\end{eqnarray}
\begin{eqnarray}
\Delta_{\vec{\eta}_{m},\vec{\chi}_n}&=&\sum^{m}_{i=1}K^{(\sigma_{m})}_{\eta_i}-\sum^n_{i=1}K^{(\sigma_n)}_{\chi_i},
\end{eqnarray}
\begin{eqnarray}\label{hfunction}
&& h_{\vec{\eta}_{m+1},\vec{\chi}_m}=\nonumber\\
&&\frac{\prod_{i>i'}(e^{-iK^{(\sigma_m)}_{\chi_i}}-e^{-iK^{(\sigma_m)}_{\chi_{i'}}})\prod_{j>j'}(e^{iK^{(\sigma_{m+1})}_{\eta_j}}-e^{iK^{(\sigma_{m+1})}_{\eta_{j'}}})}{\prod^m_{i=1}\prod^{m+1}_{j=1}(1-e^{-i(K^{(\sigma_{m+1})}_{\eta_j}-K^{(\sigma_m)}_{\chi_i})})},\nonumber\\
\end{eqnarray}
and
\begin{eqnarray}
\bar{h}_{\vec{\chi}_n,\vec{\chi}'_n}&=&\sum_{\vec{\eta}_{n+1}}h_{\vec{\eta}_{n+1},\vec{\chi}_n}h^*_{\vec{\eta}_{n+1},\vec{\chi}'_n}.
\end{eqnarray}
\par In the following, the real-time dynamics of the system will be simulated by a direct diagonalization of $H^{(\mathcal{N})}_{\mathrm{F-D}}$. We would like to mention that a slightly modified formulation of the above-mentioned spin-operator matrix element method also allows us deal with inhomogeneous light-matter interactions $\{g_j\}$~\cite{XXZCSM}, though it is more time-consuming to simulate the real-time dynamics of inhomogeneous systems due to the lack of translational invariance of the composite system.
\subsection{Energy spectrum of the molecular aggregate}
\par Before ending this section, we finally look at the energy spectrum of the pure molecular aggregate in the absence of the cavity. For $A\neq 0$, the eigenenergies and eigenstates of $H_{\mathrm{m}}$ can in principle be obtained by the Bethe ansatz method~\cite{Bethe}. Here, we use the exact diagonalization method based on the foregoing spin-operator matrix elements to solve an aggregate of $N= 16$ monomers.
\begin{figure}
\includegraphics[width=.52\textwidth]{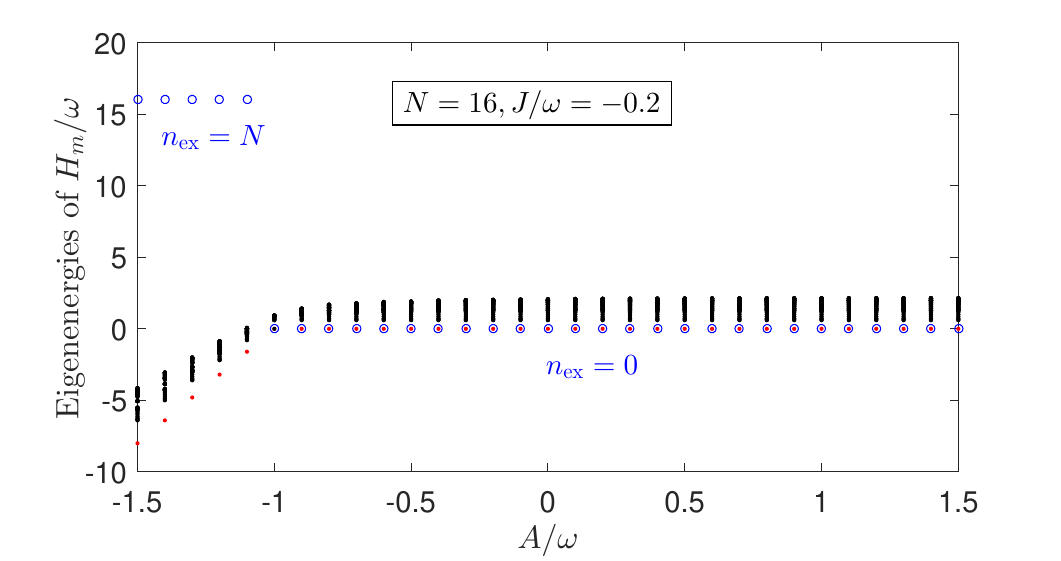}
\caption{Evolution of the lowest 100 energy levels of the molecular Hamiltonian $H_{\mathrm{m}}/\omega$ with $N=16$ and $J/\omega=-0.2$ when the exciton-exciton interaction $A/\omega$ is varied from $-1.5$ to $1.5$. The ground-state energies are highlighted in red. The blue open circles indicate the number of excitations upon the ground state. A level crossing takes place at $A/\omega=-1.0$.}
\label{Fig3}
\end{figure}
\par It is easy to see from Eq.~(\ref{XXZ}) that $|\downarrow\ldots\downarrow\rangle$ and $|\uparrow\ldots\uparrow\rangle$ are two obvious eigenstates of $H_{\mathrm{m}}$ with eigenenergies $0$ and $N(A+\omega)$, respectively. We thus expect that a ground-state level crossing occurs at $A/\omega=-1.0$ for not too large $|J|/\omega$. Figure~\ref{Fig3} shows the lowest 100 energy levels of $H_{\mathrm{m}}/\omega$ as functions of the exciton-exciton interaction $A/\omega$ for a molecular aggregate containing $N=16$ monomers (we choose $J/\omega=-0.2$, corresponding to the case of a J-aggregate). The ground-state level is highlighted in red and the blue circles represent the number of excitations $n_{\mathrm{ex}}$ in the aggregate upon the ground state. It can be seen that a level crossing does take place at $A/\omega=-1.0$ for the chosen parameters.
\par Below we focus on the case of $A/\omega>-1.0$ for which the ground state of $H_{\mathrm{m}}$ is simply the vacuum state $|G\rangle=|\mathrm{vac}\rangle=|\downarrow\ldots\downarrow\rangle$ with zero energy. In this case, the numerical simulation of the dynamics can be simplified by noting that the ground state $|G\rangle$ carries zero crystal momentum. Due to the translational invariance of the total system, the time evolution takes place within this zero-momentum subspace, as manifested in the $\delta$-functions appearing in Eqs.~(\ref{F}) and (\ref{G}).
\section{Stored energy and charging power}\label{SecIII}
\begin{figure}
\includegraphics[width=.53\textwidth]{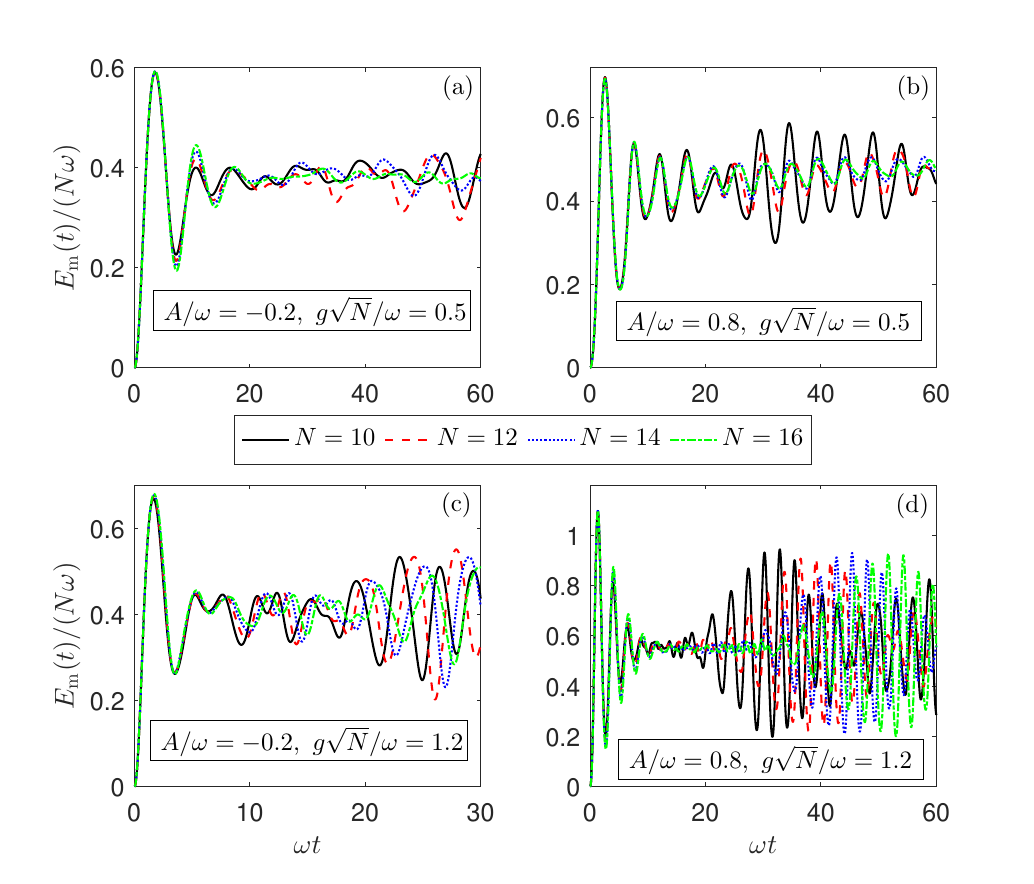}
\caption{Evolution of the stored energy density $E_{\mathrm{m}}(t)/(N\omega)$ for (a) $A/\omega=-0.2$ and $g\sqrt{N}/\omega=0.5$,  (b) $A/\omega=0.8$ and $g\sqrt{N}/\omega=0.5$, (c) $A/\omega=-0.2$ and $g\sqrt{N}/\omega=1.2$, (d) $A/\omega=0.8$ and $g\sqrt{N}/\omega=1.2$. We choose $J/\omega=-0.2$ and use normalization I to keep $g\sqrt{N}/\omega$ constant for different $N$'s.}
\label{N14vs16}
\end{figure}
\par We consider the case of J-aggregates (with $J<0$) in the numerical simulations. We also allow for both negative and positive values for the exciton-exciton interaction $A$~\cite{Ezaki1994}. The cavity frequency is set to be resonant with the onsite energy of the monomers, i.e., $\omega_{\mathrm{c}}=\omega$. The cavity photon number is set to be identical to the number of monomers, i.e.,$n_{\mathrm{ph}}=N$~\cite{QB2018}.
\par We first show that under normalization I the short-time dynamics of the system is indeed universal for different $N$'s. Figure~\ref{N14vs16} shows the dynamics of the stored energy density $E_{\mathrm{m}}(t)/(N\omega)$ for several combinations of ($A/\omega,g\sqrt{N}/\omega)$ and $N=10,12,14,16$, where the normalization I is used. We see that, for all the cases considered, $E_{\mathrm{m}}(t)/(N\omega)$ is almost independent of $N$ in the early stage and is expected to faithfully capture the short-time dynamics in the thermodynamic limit $N\to\infty$. Similar size-insensitive short-time dynamical behaviors were also observed in other many-body systems~\cite{XXZCSM,Kollath,PRE2020}. However, note that $E_{\mathrm{m}}(t)/(N\omega)$ experiences intensive fluctuations (especially for smaller $N$) at late times due to the finite-size effect. It is expected that $E_{\mathrm{m}}(t)/(N\omega)$ roughly approaches a steady-state value for large enough $N$.
\par We are interested in the maximum stored energy density $E^{(\max)}_{\mathrm{m}}/(N\omega)$~\cite{QB2018}, which is defined to be the maximal value of $E_{\mathrm{m}}(t)/(N\omega)$ in the time interval $\omega t\in[0,100]$. Usually, $E^{(\max)}_{\mathrm{m}}/(N\omega)$ shows up as the first peak of $E_{\mathrm{m}}(t)/(N\omega)$ (Fig.~\ref{N14vs16}). However, for very weak exciton-cavity couplings it is possible that $E^{(\max)}_{\mathrm{m}}/(N\omega)$ corresponds to the second peak of $E_{\mathrm{m}}(t)/(N\omega)$. Since the first maximum of $E_{\mathrm{m}}(t)/(N\omega)$ converges well for different $N$'s, we mainly choose $N=14$ in numerical simulations involving fixed $N$, but will extend the calculations up to $N=18$ when discussing the scaling behaviors with varying $N$. There are also cases in which $E_{\mathrm{m}}(t)/(N\omega)$ blows up and does not exhibit any maximum in a finite period of time. We also avoid considering these parameter regimes.
\par
\begin{figure}
\includegraphics[width=.53\textwidth]{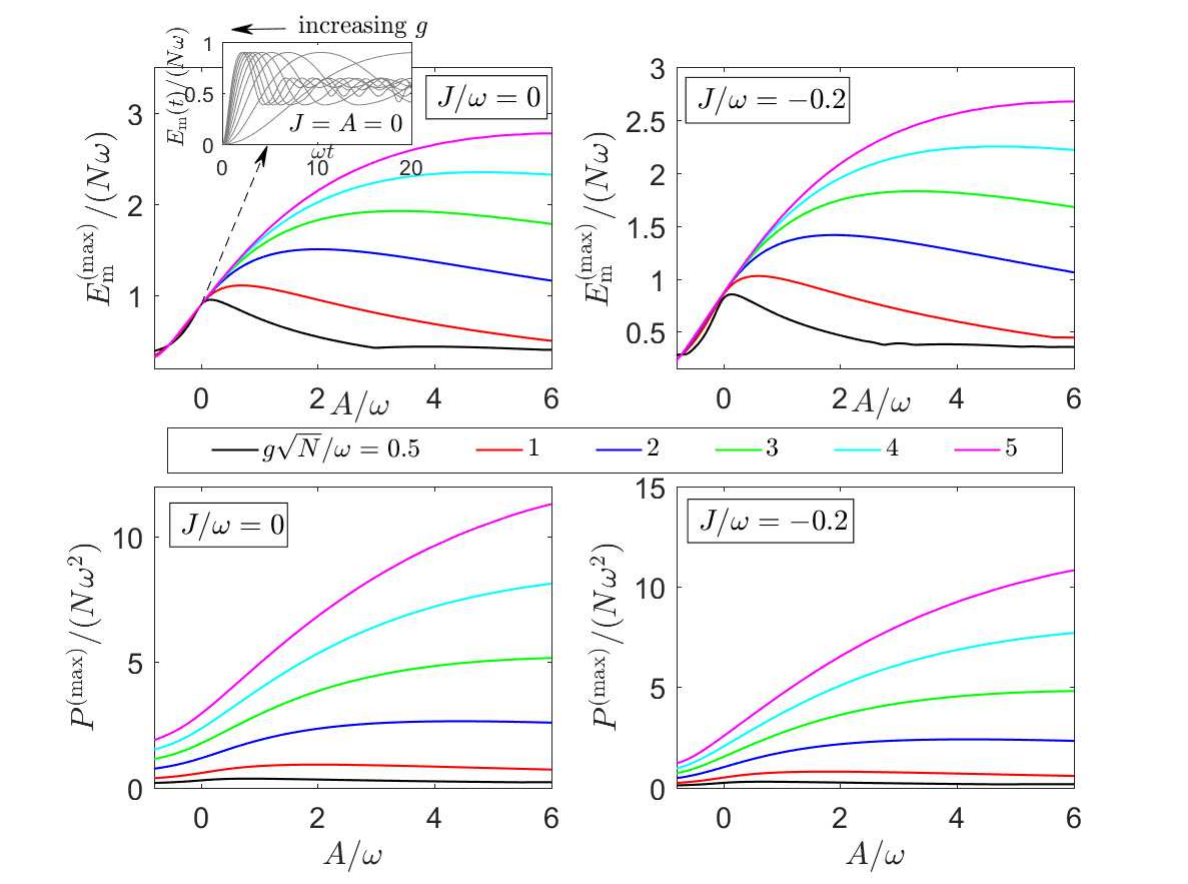}
\caption{Upper panels: Maximum stored energy density $E^{(\max)}_{\mathrm{m}}/(N\omega)$ as a function of the exciton-exciton interaction $A/\omega$. The inset shows the dynamics of $E_{\mathrm{m}}(t)/(N\omega)$ at the noninteracting point $J=A=0$. Lower panels: Maximum charging power density $P^{(\max)}/(N\omega^2)$ as a function of the exciton-exciton interaction $A/\omega$. We choose $N=14$ in the numerical simulations.}
\label{Fig4}
\end{figure}

 The upper two panels of Fig.~\ref{Fig4} show the maximum stored energy density $E^{(\max)}_{\mathrm{m}}/(N\omega)$ as a function of the exciton-exciton coupling strength $A/\omega$ from the weak to strong exciton-cavity coupling regimes. For $A/\omega\lesssim0$, $E^{(\max)}_{\mathrm{m}}/(N\omega)$ is almost independent of the exciton-cavity coupling; while for fixed $A/\omega \gtrsim0$ the stored energy density increases with increasing $g\sqrt{N}/\omega$. In particular, in the noninteracting case with $J=A=0$~\cite{QB2018}, $E^{(\max)}_{\mathrm{m}}/(N\omega)$ is almost independent of $g\sqrt{N}/\omega$ [inset of Fig.~\ref{Fig4}(a)], which is believed to be due to the RWA nature of the exciton-cavity coupling. As expected, the time at which this first maximum is reached decreases with increasing $g\sqrt{N}/\omega$. Interestingly, for both $J/\omega=0$ and $-0.2$ and for each value of $g\sqrt{N}/\omega$ considered, we observe that $E^{(\max)}_{\mathrm{m}}/(N\omega)$ exhibits a maximum at some $A_{\max,E}(g)$ that increases with increasing $g\sqrt{N}/\omega$.
\par The existence of $A_{\max,E}(g)$ can be qualitatively understood through a perturbative analysis for small $g\sqrt{N}/\omega$. Note that the first-order transition takes place between the initial ground state $|\mathrm{vac}\rangle$ and the zero-momentum one-exciton state $|\xi\rangle=\frac{1}{\sqrt{N}}\sum^N_{j=1}S^+_j|\mathrm{vac}\rangle$, and hence is independent of $A/\omega$. Thus, the lowest-order transition involving $A/\omega$ occurs between the ground state and the $N/2$ zero-momentum two-exciton states $|\phi_\alpha\rangle~(\alpha=1,2,\ldots,N/2)$ satisfying $(H_{\mathrm{m}}/\omega-2)|\phi_\alpha\rangle=\mathcal{E}_\alpha/\omega |\phi_\alpha\rangle$, where $\mathcal{E}_1\leq\mathcal{E}_2\leq\cdots\leq\mathcal{E}_{\frac{N}{2}}$ are the $A$-dependent part of the two-exciton excitation energies given by the eigenvalues of the $\frac{N}{2}\times\frac{N}{2}$ matrix~\cite{Twomagnon}
\begin{eqnarray}\label{h2}
h_2=\left(
      \begin{array}{cccccccc}
       A &   2J &   & &    &    &   &   \\
          2J & 0 &  2J &  &   &   &   &   \\
          &  2J & 0 &  2J  & &   &   &   \\
          &   &  2J &  0 &  &    &   &   \\
           &   &    &  & \ddots&   &   &   \\
          &   &   &     & &0 &  2J &   \\
          &   &   &    &  &2J & 0 &  2 \sqrt{2}J \\
          &   &   &    & &  &   2 \sqrt{2}J & 0 \\
      \end{array}
    \right),\nonumber
\end{eqnarray}
which is the matrix representation of $H_{\mathrm{m}}-2\omega$ in the ordered Bloch basis~\cite{Twomagnon}
\begin{eqnarray}
|\xi_r\rangle&=&\frac{1}{\sqrt{N}}\sum^{N}_{j=1}S^+_jS^+_{j+r}|\mathrm{vac}\rangle~(1\leq r<N/2),\nonumber\\
|\xi_{\frac{N}{2}}\rangle&=&\sqrt{\frac{2}{N}}\sum^{\frac{N}{2}}_{j=1}S^+_jS^+_{j+\frac{N}{2}}|\mathrm{vac}\rangle.
\end{eqnarray}
The eigenstate $|\phi_\alpha\rangle$ can be expanded in terms of the Bloch states as $|\phi_\alpha\rangle=\sum^{\frac{N}{2}}_{r=1}V^{(\alpha)}_r|\xi_r\rangle$ with $V^{(\alpha)}$ the eigenvector of $h_2$.
\par From time-dependent perturbation theory, the second-order transition amplitude from $|\mathrm{vac}\rangle$ to $|\phi_\alpha\rangle$ is
\begin{eqnarray}
&&T_{\mathrm{vac}\to\phi_\alpha}(t)=-g^2\sqrt{N(N-1)}\int^t_0dt_1\int^{t_1}_0dt_2 \nonumber\\
&&e^{i(\mathcal{E}_\alpha-2J)t_1}e^{i2Jt_2}\langle\phi_\alpha|\sum_jS^+_j|\xi\rangle\langle\xi|\sum_jS^+_j|\mathrm{vac}\rangle.
\end{eqnarray}
It is easy to see that $\langle\xi|\sum_jS^+_j|\mathrm{vac}\rangle=\sqrt{N}$ and the one-exciton to two-exciton transition amplitude $S_{\xi\to\phi_\alpha}\equiv\langle\phi_\alpha|\sum_jS^+_j|\xi\rangle$ can be calculated as $S_{\xi\to\phi_\alpha}=\sqrt{2}V^{(\alpha)*}_{\frac{N}{2}}+2\sum^{\frac{N}{2}-1}_{r=1}V^{(\alpha)*}_{r}$~\cite{Twomagnon}. We thus obtain the transition probability
\begin{eqnarray}\label{P2exciton}
&&P_{\mathrm{vac}\to\phi_\alpha}(t)=\frac{g^4N^2(N-1)}{4J^2\mathcal{E}^2_\alpha(\mathcal{E}_\alpha-2J)^2}|S_{\xi\to\phi_\alpha}|^2\nonumber\\
&&~[(\mathcal{E}_\alpha-2J)^2+4J^2+ \mathcal{E}_\alpha^2 +4J(\mathcal{E}_\alpha-2J)\cos \mathcal{E}_\alpha t\nonumber\\
&&-2\mathcal{E}_\alpha(\mathcal{E}_\alpha-2J)\cos 2Jt -4J\mathcal{E}_\alpha\cos (\mathcal{E}_\alpha-2J)t].
\end{eqnarray}
The stored energy through exciting the two-exciton states is then
\begin{eqnarray}\label{E2exciton}
E^{(2-\mathrm{exciton})}_{\mathrm{m}}(t)&=&\sum^{\frac{N}{2}}_{\alpha=1}P_{\mathrm{vac}\to\phi_\alpha}(t)(2\omega+\mathcal{E}_\alpha).
\end{eqnarray}
\begin{figure}
\includegraphics[width=.53\textwidth]{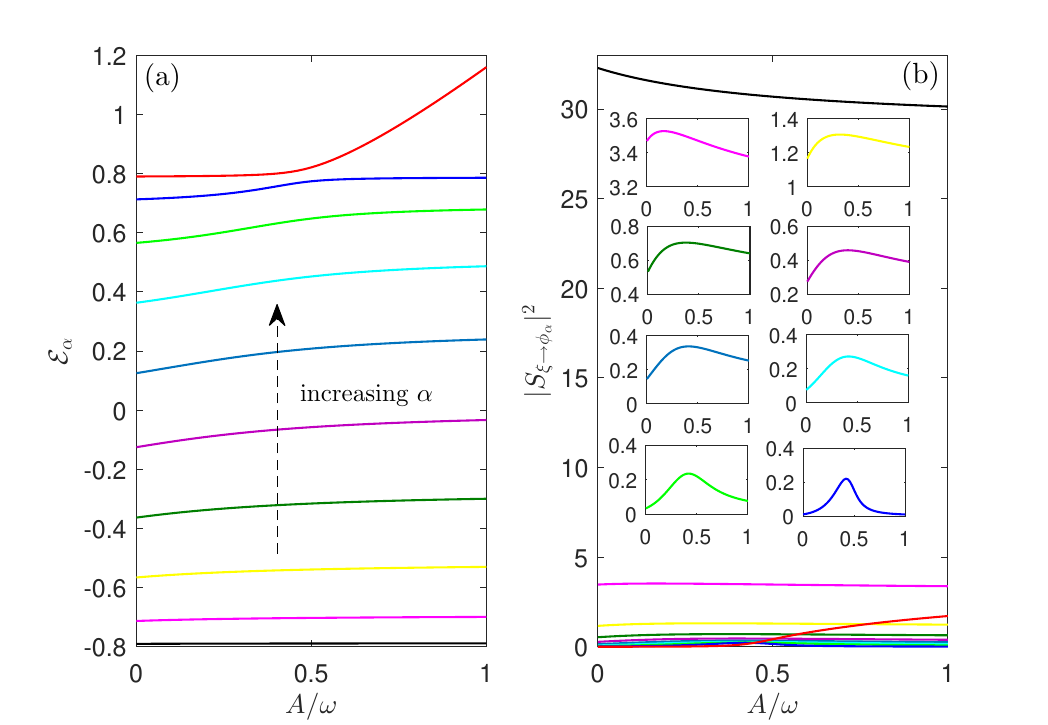}
\caption{(a) The two-exciton excitation energy $\mathcal{E}_\alpha$ as a function of $A/\omega$ for a molecular aggregate of $N=20$ monomers. (b) The corresponding one-exciton to two-exciton probability $|S_{\xi\to\phi_\alpha}|^2$. The insets show the individual plots of $|S_{\xi\to\phi_\alpha}|^2$ for $\alpha=2,3,\ldots,9$. Other parameters: $J/\omega=-0.2$ and $g\sqrt{N}/\omega=0.3$.}
\label{Fig4a}
\end{figure}
\par Although it is still not obvious to get the first maximum of $E^{(2-\mathrm{exciton})}_{\mathrm{m}}(t)$ from Eqs.~(\ref{P2exciton}) and (\ref{E2exciton}), we see that the behavior of $E^{(2-\mathrm{exciton})}_{\mathrm{m}}(t)$ may roughly depend on two factors, i.e., the one-exciton to two-exciton probability $|S_{\xi\to\phi_\alpha}|^2$ and the energy $\mathcal{E}_\alpha$. Figure~\ref{Fig4a} show both of the two quantities for $N=20$, $J/\omega=-0.2$, and $g\sqrt{N}/\omega=0.3$. The lowest nine two-exciton excitation energies $\mathcal{E}_{1},\ldots,\mathcal{E}_{9}$ increase slightly with increasing $A/\omega$, while the highest one $\mathcal{E}_{10}$ increases more rapidly for $A/\omega\geq 0.4$ due to the formation of two-exciton bound states in this regime~\cite{SpanoCPL1995,Twomagnon}. On the other hand, from Fig.~\ref{Fig4a}(b) we can see that the associated $|S_{\xi\to\phi_\alpha}|^2$ for $\alpha=1$ ($\alpha=10$) decreases (increases) monotonically as $A/\omega$ increases. As a result, it is expected that the local maximum of $E^{(2-\mathrm{exciton})}_{\mathrm{m}}(t)$ does not arise the lowest and highest two branches with $\alpha=1$ and $\alpha=10$, though the contribution from the transition to the bound state $|\phi_{10}\rangle$ becomes more dominated for $A/\omega>0.4$. However, we observe that the $|S_{\xi\to\phi_\alpha}|^2$ for $\alpha=2,\ldots,9$ all exhibit local maxima at certain $A/\omega$, which is believed to be the reason for the existence of $A_{\max,E}(g)$.
\par In the lower two panels of Fig.~\ref{Fig4} we present the corresponding maximum charging power density $P^{(\max)}/(N\omega^2)=\max_{\omega t\in[0,100]}[P(t)/(N\omega^2)]$. For fixed $A/\omega\geq -0.8$, it can be seen that $P^{(\max)}/(N\omega^2)$ increases when $g\sqrt{N}/\omega$ is increased. For fixed $g\sqrt{N}/\omega$, we also observe an optimal $A_{\max,P}(g)$ at which $P^{(\max)}/(N\omega^2)$ reaches a maximum. In general, we have $A_{\max,P}(g)>A_{\max,E}(g)$. The above results show that certain exciton-exciton interactions can facilitate the absorbtion of energy from the cavity photons.
\par
\begin{figure}
\includegraphics[width=.53\textwidth]{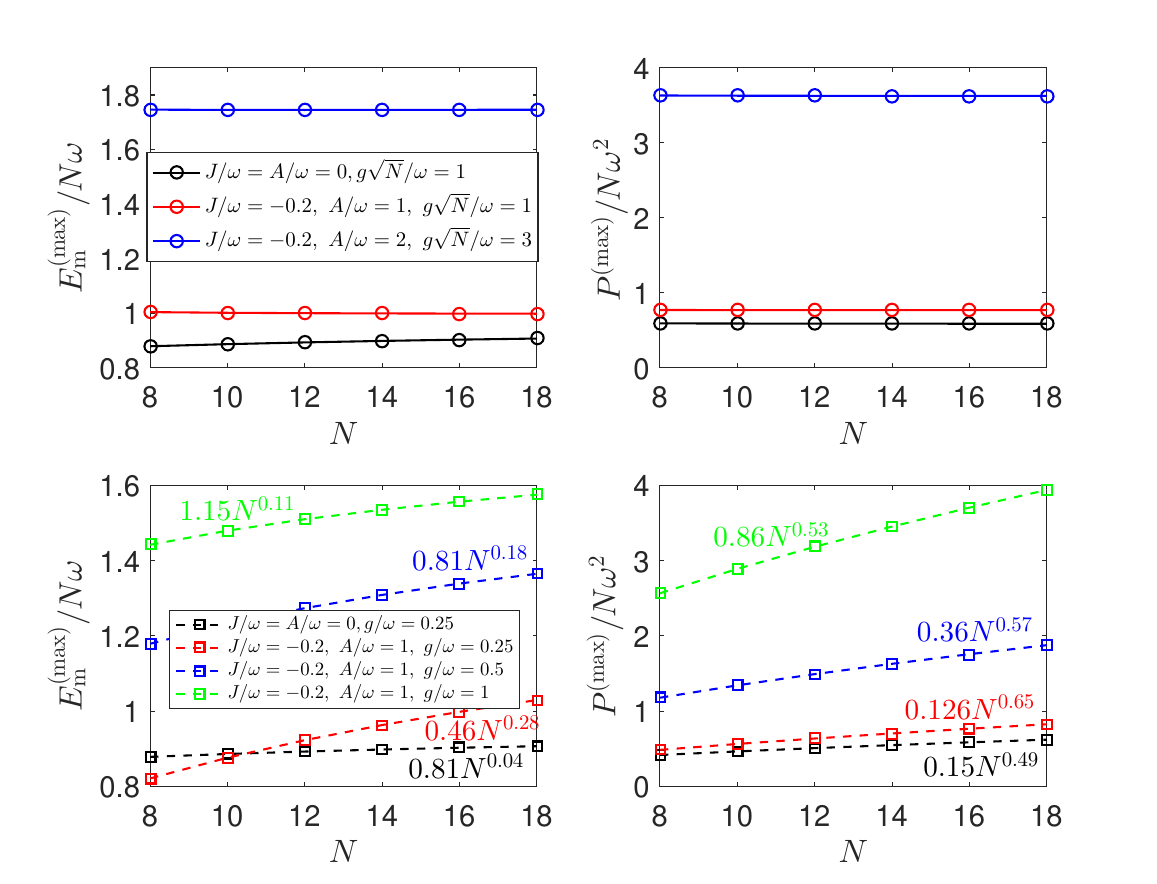}
\caption{Scaling behaviors of the maximum stored energy density $E^{(\max)}_{\mathrm{m}}/(N\omega)$ and maximum power density $P^{(\max)}/(N\omega^2)$ with varying $N$. Upper panels: normalization I, lower panels: normalization II.}
\label{Fig5}
\end{figure}

 We next discuss the scalings of $E^{(\max)}_{\mathrm{m}}/(N\omega)$ and $P^{(\max)}/(N\omega^2)$ with varying $N$. We consider both types of normalizations for the exciton-cavity coupling. The upper panels of Fig.~\ref{Fig5} show the scaling of $E^{(\max)}_{\mathrm{m}}/(N\omega)$ and $P^{(\max)}/(N\omega^2)$ under the normalization I. For three different sets of parameter with zero or finite intra-battery interactions, we see that both of the two quantities are almost size-independent, indicating that the maximum stored energy and charging power are indeed extensive quantities. In this sense, there is no quantum advantage for normalization I, despite of the fact that finite exciton-exciton interactions can enhance the maximum stored energy and charging power.
\par The lower two panels of Fig.~\ref{Fig5} show the corresponding scalings for normalization II. For the Dicke QB with $J/\omega=A/\omega=0$ and $g/\omega=0.25$, we find that $E^{(\max)}_{\mathrm{m}}/(N\omega)\approx 0.81 N^{0.04}$ and $P^{(\max)}/(N\omega^2)\approx 0.15 N^{0.49}$ (black dashed line), which are consistent with the results in Ref.~\cite{QB2018} (though the RWA is used here). Interestingly, when finite exciton hopping $J/\omega=-0.2$ and exciton-exciton interaction $A/\omega=1$ are introduced, the scalings become $E^{(\max)}_{\mathrm{m}}/(N\omega)\approx 0.46 N^{0.28}$ and $P^{(\max)}/(N\omega^2)\approx 0.126 N^{0.65}$ (red dashed line), showing that the charging process of an organic QB exhibits quantum advantages compared with the Dicke QB. However, the scaling exponents decreases with increasing $g/\omega$ (blue and green dashes lines). Thus, although for fixed $N$ the absolute values of $E^{(\max)}_{\mathrm{m}}/(N\omega)$ and $P^{(\max)}/(N\omega^2)$ increases with increasing $g/\omega$, it seems that weak exciton-cavity coupling is more favorable in enhancing the quantum advantage.
\section{Conclusions}\label{SecV}
\label{sec-final}
\par In this work, we study the charging process of a quantum battery consisting of an organic molecular aggregate and a coupled single-mode cavity, which is termed as an organic QB. The organic QB can be viewed as an extension of the Dicke QB proposed in Ref.~\cite{QB2018} to the case of finite intra-battery interactions. In contrast to several previous studies~\cite{QB2018,Dou,Zhang}, the total angular momentum of the battery part is not conserved, which causes difficulties in the simulation of dynamics of the system, especially for large battery sizes. With the help of the spin-operator matrix elements method~\cite{Wu2018} and the translational invariance of the system, we are able to obtain exact dynamics of organic QBs containing $N\leq 18$ monomers.
\par We consider two types of normalizations of the exciton-cavity coupling when $N$ is varied, i.e., one with the cavity length changing to keep the monomer density constant (type I) and the other with the cavity unchanged (type II). Numerical simulations of the dynamics of the stored energy and the charging power reveal that under normalization I both the maximum stored energy density and the maximum charging power density are extensive quantities, i.e., they do not show quantum advantages. In contrast, when normalization II adopted, we observe improved quantum advantages in the two quantities compared with the Dicke QB. In addition, we find that for fixed exciton-cavity coupling there always exists optimal exciton-exciton interactions that maximize the two quantities, which is qualitatively explained through a perturbative analysis up to two excitons for weak exciton-cavity couplings. Our work may stimulate further studies on quantum batteries involving organic materials.

\noindent{\bf Acknowledgements:}
This work was supported by the National Key Research and Development Program of China under Grant No. 2021YFA1400803.

\appendix

\end{document}